# Surgical navigation systems based on augmented reality technologies

Vladimir Ivanov[1*], Anton Krivtsov[1], Sergey Strelkov[1], Dmitry Gulyaev[2], Denis Godanyuk[2], Nikolay Kalakutsky[3], Artyom Pavlov[3], Marina Petropavloskaya[3], Alexander Smirnov[3], Andrew Yaremenko[3]

[1] Peter the Great St. Petersburg Polytechnic University, St. Petersburg, Russian Federation

[2] Almazov National Medical Research Centre, St. Petersburg, Russian Federation

[3] Pavlov First St. Petersburg State Medical University, St. Petersburg, Russian Federation

*Corresponding author

**Author emails:**
voliva@rambler.ru (Vladimir Ivanov); anton.krivtsov@spbstu.ru (Anton Krivtsov); sergin3d2d@gmail.com (Sergey Strelkov); spb.gda@yandex.ru (Dmitry Gulyaev); godanyuk@gmail.com (Denis Godanyuk); kalakutsky@yandex.ru (Nikolay Kalakutsky); tlocke@bk.ru (Artyom Pavlov); m.petropavlovskaya@mail.ru (Marina Petropavlovskaya); smirnov-1959@yandex.ru (Alexander Smirnov); yaremenkoai@spb-gmu.ru (Andrew Yaremenko)

## Abstract

This study considers modern surgical navigation systems based on augmented reality technologies. Augmented reality glasses are used to construct holograms of the patient's organs from MRI and CT data, subsequently transmitted to the glasses. This, in addition to seeing the actual patient, the surgeon gains visualization inside the patient's body (bones, soft tissues, blood vessels, etc.). The solutions developed at Peter the Great St. Petersburg Polytechnic University allow reducing the invasiveness of the procedure and preserving healthy tissues. This also improves the navigation process, making it easier to estimate the location and size of the tumor to be removed.

We describe the application of developed systems to different types of surgical operations (removal of a malignant brain tumor, removal of a cyst of the cervical spine). We consider the specifics of novel navigation systems designed for anesthesia, for endoscopic operations. Furthermore, we discuss the construction of novel visualization systems for ultrasound machines. Our findings indicate that the technologies proposed show potential for telemedicine.

## Method

We have developed software that not only loads and positions the 3D model relative to the marker but also offers an additional user interface for interaction during the surgical procedure. The interface based on gestures and virtual buttons makes it possible for the surgeon to use the extensive range of functions incorporated in the glasses without touching non-sterile objects.

We have constructed a system detecting and tracking the observer's position via the HTC Vive positioning system using two trackers, one attached to the glasses, and the other to the calibration pointer.

The advantage of this approach compared to existing optical navigation systems is the tracked space of 6 x 6 meters, capturing almost the entire operating room. The system can work with 4 base stations; tracking is possible even if the sensor is within the field of view of only one of the stations.

We have developed a versatile headset which will be adjusted to the individual anatomy of each patient. The headset is a flexible frame that is attached to the head at three main points: on the bridge of the nose and in each ear canal. The frame is shaped to provide maximum access to the brain or maxillofacial area during different types of surgeries. The adjustable frame is intended to be attached to the patient's head before the CT scan and re-attached for the surgery in the exact position that it had during the CT scan. For this purpose, the frame is equipped with an adjustment control which can be regulated with a certain step and reset to the same position; additionally, adjustable screws are provided to fix the ear tips in the ear canal. Metal markers are also installed in several points of the frame, which are enhanced with contrast on CT and allow automatically detecting the position of the marker relative to the patient's head. The marker itself, in turn, is removable and does not have to be installed during CT.

**Results**

**Clinical Case 1 (surgery to remove a brain tumor)**
A 54-year-old patient was hospitalized with progressive glioblastoma in the left frontal lobe.

Brain MRI revealed an intracerebral tumor of the left frontal lobe with mass effect and indistinct margins; contrast uptake was moderately inhomogeneous, the part of the tumor enhanced with the contrast agent had a size of 7 x 6 x 6 cm, the adjacent structures of the brain were compressed, the median structures were displaced to the right.

As the tumor progressed, the patient developed symptoms of intracranial hypertension; in view of this, the surgical team at the Almazov National Medical Research Centre performed the world's first operation to remove the brain tumor using mixed reality glasses: Repeat left frontotemporal craniotomy, microsurgical removal of the growing tumor of the left frontal lobe with neurophysiological control and multimodal navigation (Medtronic Stealts Station S7 + Hololens mixed reality).

Mixed reality was used for preoperative planning, intraoperative marking of the operative field, and control of surgical aggressiveness.

After the patient was positioned, the frame was mounted; the tumor was then localized and surgical access to it was planned (Fig. 1).

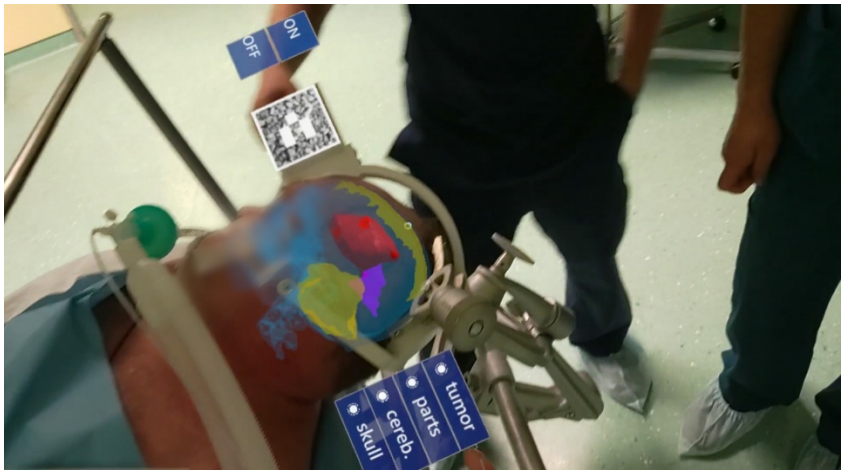

*Figure 1: View of tumor and auxiliary anatomical structures through augmented reality glasses.*

Using mixed reality at the planning stage allowed achieving optimal access and avoiding unnecessary tissue dissection.

After the dura mater was approached via mixed reality, the surgical site was marked and labeled to estimate the intended durotomy length. Resection radicality was constantly monitored during the surgery. Intraoperative monitoring data indicated that the tumor was completely removed, which was confirmed by the data of a postoperative brain CT.

**Clinical Case 2 (surgery to remove a cyst in the cervical spine)**

The described approach was implemented during a surgical intervention in a female patient with a midline neck cyst. This operation was performed by the surgical team at the Pavlov First St. Petersburg State Medical University; it was the first surgery of this class carried out in Russia using augmented reality technologies.

It was decided to use the visualization system based on mixed reality glasses during the operation to track the exact boundaries and syntopic characteristics of the location of the cyst and fistula; this way, the spatial relationship and localization of the cyst relative to the neck organs could be determined as accurately as possible accounting for the patient-specific anatomy. Simultaneous intraoperative neuromonitoring of nerve integrity was performed via the Medtronic NIM-neuro 3.0 system.

Because the operative field was localized in the cervical spine, the patient's position during the surgery had to repeat the exact position during the MSCT scan to position the hologram precisely. A mask immobilizing the head and shoulder girdle in a certain position, allowing to reproduce this position during a future operation, was used for this purpose (Fig. 2).

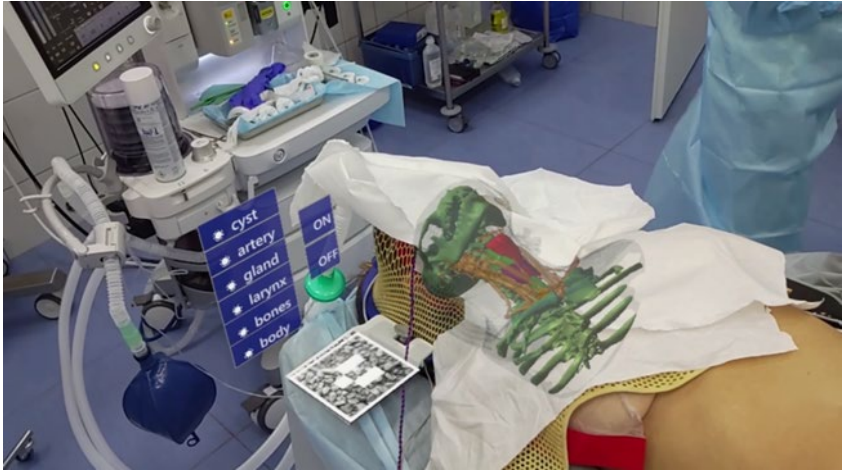
*Figure 2: View of hologram through Hololens 2*

Analyzing the clinical experience accumulated using the mixed reality technology in surgery, we can conclude that this technique provided an improved visualization of the real situation within the operative field, yielding a clearer intraoperative picture on the spatial localization of malignant neck tumors relative to vital structures (vessels, nerves, hollow organs) to complement the data obtained at the preoperative stage during ultrasound, MSCT and MRI studies.

**Clinical Case 3 (peculiarities of endoscopic operations)**
As we can see, the experience of using augmented reality technologies for different types of surgical interventions has been largely positive; let us now consider another type of surgery, namely, the specifics and prospects of endoscopic operations.

Endoscopy has come a long way from an exclusively diagnostic technique with the sole task of detecting the extent of disease to an independent discipline using flexible tools inserted through the body's natural orifices to treat conditions that could previously only be solved by surgical methods. A clear benefit of flexible endoscopy is its low invasiveness: the endoscopic tools are inserted into the patient's body through the natural orifices (e.g., mouth, or anal canal); moreover, the endoscope can take the shape of the structures where the procedure is performed. The inserted part of the endoscope can combine multiple functions, serving as a camera transmitting images, a means for delivering instruments for performing manipulations, and a tube supplying gases and liquids to generate a working environment. Natural orifice transluminal endoscopic surgery (NOTES) performed in cavities and organs accessed through the mucous membranes of the oral cavity, esophagus, rectum seems to be a promising direction in flexible endoscopy. Transesophageal access to the mediastinum is an advantageous technique that is fairly well understood by now thanks to tunnel endoscopic surgeries in the esophagus (POEM, STER), minimizing the access trauma and delivering the necessary tools to the operative field within minutes from the start of the surgery. The issue of navigation remains unresolved because minimizing access implies that the tunnel formed should lead precisely to the area of interest (the hit-to-kill principle);

otherwise, the operating surgeon is forced to spend a large amount of time searching for the target mediastinum site, which may carry an increased risk of complications. Since intraoperative navigation remains a challenge at present, the level of the tunnel and the direction in which it is formed are chosen intuitively, based on the experience of the operating surgeon and the data of presurgical examination. Direct intraoperative navigation, where not only the area of interest but also the mediastinal structures would be displayed on the endoscopic surgeon's screen (Figure 3), can greatly simplify the surgery, speed it up, make it safer, and, accordingly, improve its reproducibility, turning it into a standard surgical procedure accessible to many surgeons.

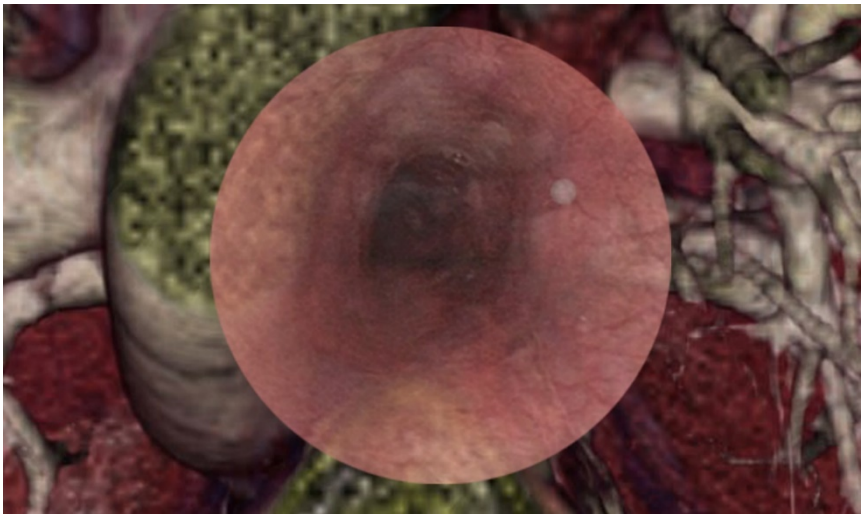

*Figure 3: Endoscopic image combined with augmented reality*

The augmented reality technologies described also have promising applications in anesthesiology. Being able to see vessels through the glasses and visualize the position of the syringe needle by ultrasound data makes it possible to build a completely new technology for these types of surgeries.

Finally, combining augmented reality technologies with telemedicine technologies means that operations can be performed remotely by surgical team members from different locations. Surgeons outside the OR can detect/outline the risk zones directly on their screens, transmitting this information to the operating surgeon's glasses.

**Conclusions**

Surgical navigation systems generating virtual 3D images of the operative field combined with the intraoperative picture can revolutionize the concept of preoperative planning and surgical navigation, prompting a shift towards intelligent and intuitive assistance. Additionally, the quality of surgical treatment can be improved, preparing the grounds for extremely sophisticated, often unparalleled surgical interventions.

We believe that integrating such systems into everyday surgical practices will have not produce positive clinical and economic outcomes but will be greatly beneficial for the emotional wellbeing of both the recovered patient and the surgeon providing the medical care.